\documentclass[12pt]{article}
  \usepackage{amsthm,amsfonts}
  \usepackage{amsmath}
\newcommand{\bea}   {\begin{eqnarray}}
\newcommand{\eea}   {\end{eqnarray}}
\usepackage{subeqnarray}
\usepackage[brazilian]{babel}
\usepackage[latin1]{inputenc}
\begin{document}
\renewcommand{\thefootnote}{\fnsymbol{footnote}}
\date{}

\begin{center}
{\bf On the Dirac Monopole Mass Scale}
\end{center}

\begin{center}
{\it F. Caruso\footnote{On leave of absence from the Instituto de F\'{\i}sica da Universidade do Estado
do Rio de Janeiro, RJ, Brasil.}}\\
\vspace*{0.2cm}
Istituto di Fisica della Universit\`{a} di Torino, \\
Istituto Nazionale di Fisica Nucleare, Sezione di Torino, Italy
\end{center}

\vspace*{3mm}
\noindent

\vspace*{3mm}
\noindent
{\bf Abstract:} It is shown, by a semi-classical argument, that the Dirac
charge quantization is still valid in the (classical) Born-Infeld electromagnetic
theory. Then it is possible to calculate Dirac's monopole
mass in the framework of this theory, which is not possible in Maxwell's
theory. The existence o f an upper limit for the field intensities in this
theory plays an important role in this proof.

\vspace*{0.2cm}

\section{Introduction}
\setcounter{footnote}{0}
\renewcommand{\thefootnote}{\arabic{footnote}}
The main motivation of the Born-infield (B.I.) electromagnetic theory \cite{um} was to avoid the divergent se1f-energy of point charges.
Maxwell's electromagnetic theory was generalized in an elegant way, by using general symmetry principles which, in the last instance, brings in a non scale-invariant field theory \cite{um}.d.

In this paper, it will be shown that the existence of a scale parameter -- an upper limit for the field \textit{intensity} -- in this generalized
electromagnetic field theory, allow us to {\it obtain}, in a straightforward way, a relation between the magnetic and e electric masses $(m_M,m_e)$ and
charges $(g,e)$ which differs from the usual one. Following the same semi-classical argument \cite{dois} that we have recently used to obtain the quantization condition for the charges of the dyons \cite{tres} in Maxwell's
theory, we can see that the Dirac quantization equation. For the magnetic and electric charges is {\it still valid} for the B.I. theory. Then, it follows that within the framework of this theory, we can compute the actual value of the monopole mass, which is not possible in Dirac's original formulation of the magnetic monopoles \cite{quatro}, based on Maxwell's theory. Indeed we like
to emphasize that in the Dirac formulation of magnetic monopoles there is no {\it prediction} for the monopole mass; the estimate of the Dirac monopole mass; is possible only if one {\it postulates} that the ``classical''  monopole radius is equal to the electron ``classical'' radius~\cite{cinco}. Thus one finds $m_M=m_eg^2/e^2=2.4$~GeV.

The arguments that lead to the relation $m_M=m_e(g^2/e^2)^{3/4}$ for
the Born-Infeld theory, without any hypothesis on the electric and magnetic radii, are presented in section~\ref{relation} (notice the change in the power).
Conclusions are presented in section~\ref{conc}.

\vspace*{0.2cm}

\section{The new relation between $m_M$ and $m_e$}\label{relation}

In the Born-Infeld Theory, the field equations (which reduce to Maxwell's equations in the limit of weak field intensities) can be
also derived from the postulate that there exists an {\it absolute field} $b(=10^{16}$~e.s.u.) which is a natural limit for the field intensity\footnote{It is not our aim in this paper to discuss the general consequences of
this interpretation, although we believe that it deserves some attention.} In this theory all the electron mass has an electromagnetic origin, which
was not a new idea in those times. The electrostatic solution with \cite{um}.d spherical symmetry is taken as a model for the classical electron,
where the electronic charge e is, actually , an integration constant ($e$ is not an {\it a priori} attribute of the particles). The energy of this
solution is equated to the rest energy of the electron giving:
\begin{equation} \label{me}
m_ec^2=1.2361(be^3)^{1/2}=1.2361\ e^2/r^{(e)}_0
\end{equation}
thus defining the classical radius $r^{(e)}$ by:
\[
r_0^{(e)}=(e/b)^{1/2}
\]
In the same way we can identify the spherically symmetric magnetostatic solution (not investigated in the Born-Infeld paper of 1934) with the magnetic monopole whose charge $g$ is another integration constant. In
this case we have the analogous equations:
\begin{equation}\label{mm}
m_Mc^2=1.2361(bg^3)^{1/2}=1.2361\ g^2/r_0^{(M)}
\end{equation}
\[
r_0^{(M)}=(g/b)^{1/2}
\]
Equating the value of $b$ taken from (\ref{me}) and (\ref{mm}), and as the $b$ parameter in the theory is unique, it follows that:
\begin{equation}\label{mm-pred}
m_M=m_e(g^2/e^2)^{3/4}
\end{equation}
which differs from the usual relation only by the power of $(g^2/e^2)$. We now find a Dirac-like relation between $g$ and $e$, thus obtaining a numerical value for $m_M$ in terms of $m_e$.

First, we remember that it was shown by B.I. \cite{um} that an elementary charge, on which an external field is acting, exactly satisfies the Lorentz's equation of motion, as it does in Maxwell theory,
whenever the external potential is essentially constant within the radius $r_0$, of the elementary charge considered. On the other hand, we know that we obtain the Dirac quantization condition $eg=n/2$ $(\hbar =c=1)$
by considering, in Maxwell theory, the semi-classical scattering of an electric charge by the magnetic field action (Lorentz force) produced
by a stationary magnetic charge, and then, computing the angular momentum variation $\Delta L$.\footnote{Indeed it is shown that Dirac's quantization condition is intimately related to the quantization of angular momentum \cite{seis}  despite the naive semi-classical approach used in this derivation~\cite{cinco}.} This quantity does not depend on the particle velocity neither on the impact parameter of the collision, $d$, being $\Delta L= 2eg$. We shall now show how the same condition $eg=n/2$ can be obtained also in the framework of B.I. theory by using the same aforementioned procedure, as the equations of motion are the same in both theories.

We consider an external ``Coulombian'' field produced by as stationary magnetic monopole charge acting on the incident electron.
Whenever the impact parameter $d$ is large enough, $d>>r_0^{(e)}$, the classical Lorentz's force law is still valid and, thus, the semi-classical argument of ref.~\cite{dois} can be applied to compute AL. Indeed, due to this
condition, the strength of the ``Coulombian'' field can be considered weak in the region within the electron radius. Here which should be an integer number by the class
With this result, we can write eq.~(\ref{mm-pred}) as
\begin{equation}
m_M=m_e(137 \ n/2)^{3/4},
\end{equation}
which gives $m_M=290$~MeV (for $n=1$)  if we use the experimental value it follows also that the two ``classical'' radii are equal:
\[
r^{(M)}_0=8.3\ \sqrt{n}\ r^{(e)}_0.
\]
Therefore the generalization of Maxwell's theory to B.I. theory does not maintain the relation $r^{(M)}=r^{(e)}$, which is usually used to estimate  $m_M$ \cite{cinco}. Besides we now obtain now the lower numerical value $m_M=0.29$~GeV instead of the value 2.4~GeV without the \textit{ad hoc} position of any relation between $r^{(M)}$ and $r^{(e)}$.

\vspace*{0.2cm}

\section{Conclusions}\label{conc}

Following the idea that one may have to modify Maxwell's theory (accepted by Dirac himself \cite{sete}), if we accept the Born-Infeld theory as a natural generalization of Maxwell's electromagnetic theory, we
obtain some improvements concerning the study of Dirac's monopoles, This theory contains two regular spherically symmetric static solutions (for
$\vec{E}$ and $\vec{H}$) which could be taken as models for the ``classical'' electron and magnetic monopole, following the original ideas of B.I.. With this
assumption, we have shown that, in the framework of this theory, it is possible to {\it calculate} the monopole mass with no a {\it priori} restriction
on the relation $r^{(M)}_0=r^{(e)}_0$ (notice that the hypothesis $r_0^{(M)}=r_0^{(e)}$ used in the standard derivation of monopole mass has indeed no logical support
in the framework of Maxwell's theory). This is, in the last analysis, a consequence of two main properties of that electromagnetic theory:
first, it is not scale-invariant (as a consequence of it non linearity), and it has a unique scale parameter $b$; second the Dirac
charge quantization between the charges $g$ and is still valid.

Finally, instead of the usual relation $m_M=m_e(g^2/e^2)$, we found $m_M=m_e(g^2/e^2)^{3/4}$. The numerical result $m_M=0.29$~GeV is one order of magnitude smaller than the usual value $m_M=2.4$~GeV. One can argue about
the discrepancy between these two values (and also about their accuracy), but it is important to stress that both estimates have a semi-classical nature and, therefore, some care must be exercised before
taking any one as an indicative value for the ``classical'' (Dirac-like) monopole mass. Although the condition $r_0^{(M)}=r_0^{(e)}$ is {\it ad hoc} in Maxwell's theory, the equivalent condition is given by B.I.'s theory. In any way, the monopole mass scale does not seem to be the reason for monopoles not having been experimentally detected. In summary, the reason why monopoles (Dirac's-like or not) are not yet found in nature is somewhat
of a mystery and remains an open question in physics.

\newpage
The author would like to thank Prof. E. Predazzi of the Istituto di Fisica Teorica della Universit\`{a} di Torino, where this work was done, for this hospitality and critical comments.
He also thanks Dr. M.O. Calv\~{a}o for stimulating discussions about the Born-Infeld theory and Prof. J. Tiomno for many useful conversations and critical comments.
This work was supported by the Conselho Nacional de Desenvolvimento Cient\'{\i}fico e Tecnol\'{o}gico, CNPq, of Brazil.


\vspace*{1cm}
\begin{center}
Resumo
\end{center}

Se mostra, por um argumento semi-cl\'{a}ssico, que a quantiza\c{c}\~{a}o
de Dirac \'{e} ainda v\'{a}lida na teoria eletromagn\'{e}tica (cl\'{a}ssica) de Born-Infeld. Pode-se ent\~{a}o calcular a massa do monopolo de Dirac dentro do
formalismo desta teoria, o que n\~{a}o \'{e} poss\'{\i}vel na teoria de Maxwell. A
exist\^{e}ncia de um limite  superior, nesta teoria, para as intensidades de
campo desempenha um papel importante nesta prova.

\begin{thebibliography}{99}
\bibitem{um}
a) M. Born, \textit{Nature} \textbf{132}, 282 (1933);  b) \textit{Proc. Roy. Soc.} \textbf{143}, 410 (1934); c) M. Born and L. Infeld, \textit{Nature} \textbf{132},
1004 (1933); d) \textit{Proc. Roy. Soc.} \textbf{144}, 425 (1934); e) \textit{ibid.} \textbf{147}, 522 (1934); f) \textit{ibid} \textbf{150}, 141 (1935).
\bibitem{dois} A.S. Goldhaber, \textit{Phys. Rev. B} \textbf{140}, 1407 (1965).
\bibitem{tres} F. Caruso ``On the origin of the charge quantization condition of the
dyons'' (in Italian), to appear in the \textit{Giornale di Fisica} (in press) [\textit{Giornale di Fisica} \textbf{XXVII} (2), 149-152 (1986)].
\bibitem{quatro} P.A.M. Dirac, \textit{Proc. R. Soc. London} \textbf{133}, 60 (1931).
\bibitem{cinco} See, for example, G. Giacomelli, ``Magnetic Monopoles'', in \textit{la Rivista
del Nuovo Cimento} \textbf{7}, n.~12 (1984).
\bibitem{seis} R.J. Crewther, ``Introduction of the theory of magnetic monopoles'', in
\textit{Proceedings of the 1982 Zuoz Spring School}, SIN Switzeland.
\bibitem{sete} ``For example, one would like to take in to account the possibility
that Maxwell's equations are not accurately valid. When one goes to
distantes very close to the charges that are producing the fields,
one may have to modify Maxwell's field theory so as to make it into
a nonlinear electrodynamics'', from P.A.M. Dirac, \textit{Lectures on Quantum
Mechanics}. (Belfer Graduate School of Science Yeshiva University,
New York, 1964), p.~2.
\end{thebibliography}
\end{document}